\documentclass[nocopyright]{cimento}

\usepackage{graphicx}

\title{Production of Resonances in Partial Chemical Equilibrium}
\author{B. Tom\'a\v{s}ik\from{ins:fjfi}\from{ins:umb}
\atque
S.~L\"ok\"os\from{ins:ifj}\from{ins:elte}\from{ins:mate}
}
\instlist{%
\inst{ins:fjfi} Fakulta jadern\'a a fyzik\'aln\v{e} in\v{z}en\'yrsk\'a, \v{C}esk\'e vysok\'e u\v{c}en\'i technick\'e v Praze, B\v{r}ehov\'a 7, 11519 Praha 1, Czech Republic
\inst{ins:umb} Fakulta pr\'irodn\'ych vied, Univerzita Mateja Bela, Tajovsk\'eho 40, 97401 Bansk\'a Bystrica, Slovakia
\inst{ins:ifj} The Henryk Niewodnicza\'nski Institute of Nuclear Physics, Polish Academy of Sciences, ul. Radzikowskiego 152, 31-342 Krak\'ow, Poland
\inst{ins:elte} E\"otv\"os Lor\'and Tudom\'anyegyetem, P\'azm\'any P\'eter s\'et\'any 1/A, 1111, Budapest, Hungary
\inst{ins:mate} Hungarian University of Agriculture and Life Sciences, KRC, M\'atrai \'ut 36, 3200, Gy\"ongy\"os, Hungary
}

\begin{document}
\maketitle

\begin{abstract}
Within the model of partial chemical equilibrium (PCE) we calculate the multiplicity ratios of selected 
unstable resonances to given stable species. We focus on those ratios that have been measured either 
in Pb+Pb collisions at the LHC or in Au+Au collisions at the top RHIC energy. The model provides an
interpretation how in an expanding hadronic fireball with decreasing temperature the final numbers 
of stable hadrons after decays of all resonances remain unchanged. Each stable species acquires its 
own chemical potential and the resonances are kept in equilibrium with them. Multiplicities of unstable 
resonances provide a test of this scenario. We observe that the ratios of $K^{*}/K$ and 
$\rho^0/\pi$ fit reasonably well into the picture of single kinetic freeze-out of the single-particle spectra, 
but the $\phi$-meson and hyperon resonances are not reproduced by this model. 
\end{abstract}


\section{Introduction}

Abundances of stable hadrons and even nuclear clusters are rather well reproduced by the Statistical Hadron Resonance Gas Model \cite{Andronic:2017pug,Andronic:2016nof}.
The model represents an  interacting hadron gas, where interactions are accounted for by the inclusion of resonances as free particles \cite{Dashen:1969ep} and even better description is encountered by including (some) interactions by using the scattering phase shifts \cite{Venugopalan:1992hy,Andronic:2018qqt}. 
The crucial point for our discussion is that the chemical freeze-out temperature inferred from hadron multiplicities is above 150~MeV for nuclear collisions at the LHC and top RHIC energies \cite{Andronic:2017pug,Andronic:2016nof}. 
Note that the matter appears to be in chemical equilibrium. 
This means that the only relevant parameters in addition to volume and temperature are chemical potentials for conserved quantum numbers, notably the baryon number. 
Only the quantum numbers of a hadron species determine their chemical potentials. 

On the other hand, a locally thermalised and expanding system is also assumed when one interprets the observed single-particle $p_t$ spectra of identified hadrons. 
There is quite a large spread in the results for the kinetic freeze-out temperature, with values from about 150~MeV \cite{Mazeliauskas:2019ifr} down to as low as 80~MeV \cite{Melo:2019mpn}.

The upper value would mean that the chemical and the kinetic freeze-out happen simultaneously. 
However, if the kinetic freeze-out temperature is low, then we can take the values of chemical and kinetic freeze-out temperatures as the initial and final temperatures of the hadron gas, respectively. 
We thus have an expanding and cooling system which still keeps its chemical composition such that after decays of all unstable resonances the abundances of stable hadrons remain equal to the values at the chemical freeze-out. 
Later on, we refer to the abundances with added contributions from resonance decays  as effective. 

The usual lore is that the inelastic processes stop after the chemical freeze-out but the elastic ones continue. 
This would certainly guarantee the desired outcome, but is not strictly necessary to ensure it.

As for the resonances, in general there are several effects that will influence their observed numbers. 
Note that  they can only be reconstructed through the measurement of their daughter particles. 
If a resonance decays early in the hadronic system and (at least one of) the daughters scatter, the invariant mass of the resonance will not be reconstructed. 
On the other hand, in a system of interacting hadrons, resonances can also be regenerated. 
The actual number of finally observed resonances results from an interplay between the loss through daughter particle scattering and the gain due to regeneration in hadronic interactions.

One of the possible scenarios for such an evolution goes under the name Partial Chemical Equilibrium (PCE) \cite{Bebie:1991ij}.
Here, the word 'equilibrium` refers to the relation between unstable resonances and the ground state hadrons. 
This is  specific assumption about the decays and regeneration of the resonances discussed above. 
Here, we report on a calculation of resonance abundances in the PCE scenario applied to different centralities of  collisions of Au+Au at top RHIC energy and Pb+Pb at LHC energy \cite{Lokos:2022jze}. 

We explain the PCE model in the next Section, show the results in Section~\ref{s:results}, and conclude in Section~\ref{s:conc}.


\section{Partial Chemical Equilibrium}
\label{s:PCE}

The PCE scenario has been introduced in \cite{Bebie:1991ij}.

\begin{figure}[t]
\centering
\includegraphics[width=0.45\textwidth]{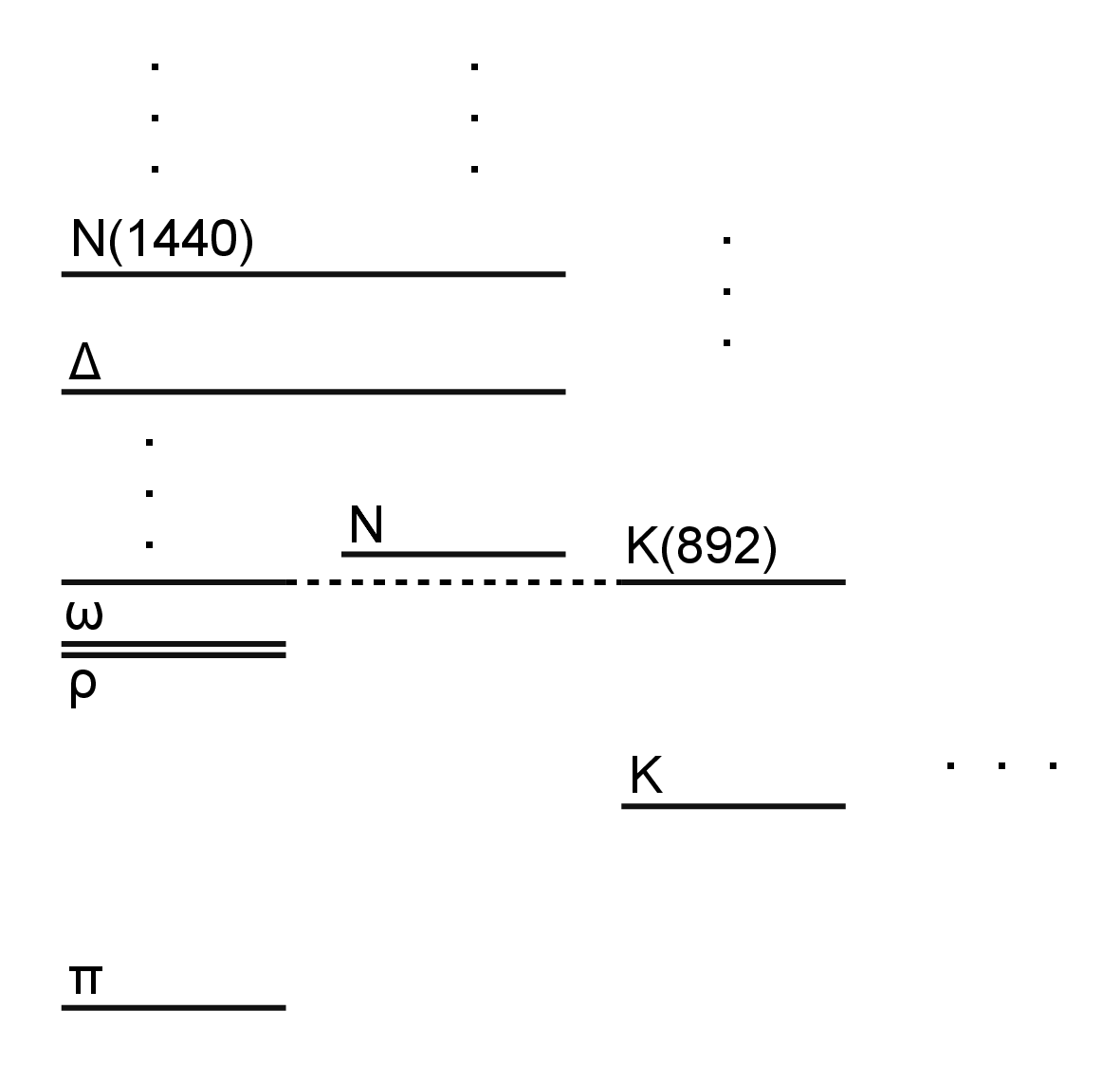}
\caption{%
A scheme of (selected) ground states of stable hadrons ($\pi$, $N$, $K$) with resonances. The vertical position of the states corresponds to their masses and the resonance states are drawn above the stable hadrons that they decay into. 
}
\label{f:states}       
\end{figure}

In order to keep the (effective) numbers of stable hadrons constant while the temperature is decreasing, there must be proper chemical potential assigned directly to each of them. Thus the overall equilibrium is lost and chemical potentials are no longer determined by the conserved quantum numbers. 
Resonances are assumed to be in equilibrium with their decay products. 
This means that their chemical potentials are given in terms of the chemical potentials of their decay products.
Figure~\ref{f:states} illustrates the scheme.
%
%
To start, let us look at resonances that decay into one sort of particles, e.g.\ $\rho$. Since the process $\rho\leftrightarrow 2\pi$ is assumed to be in equilibrium, neither of its directions should cost additional energy and therefore $\mu_\rho = 2\mu_\pi$.  For a similar reason, $\mu_\omega = 3\mu_\pi$. 
Resonances that decay into two different stable species get chemical potential by summing those of the daughter species, e.g. $\mu_{K(892)} = \mu_K + \mu_\pi$, or $\mu_\Delta = \mu_N + \mu_\pi$. 
Finally, there are resonances with multiple channels open for decay. In their case, the general formula for chemical potential of resonance $R$ can be applied
\begin{equation}
\mu_R = \sum_h p_{R\to h} \mu_h\,  ,
\end{equation}
where $p_{R\to h}$ is the average number of stable hadrons $h$ that are produced in decays of $R$, and $\mu_h$ is the chemical potential of that species. 
Here, the sum runs through all \emph{stable species} that develop their own chemical potentials. If a resonance decays into another unstable resonance, then all such decay chains are traced all the way down to the stable daughter hadrons and their average numbers appear as $p_{R\to h}$. 

The average final number of hadron species $h$ consists of directly thermally produced particles and those resulting from decays of all unstable resonances
\begin{equation}
\langle N_h^{\mathrm{eff}} \rangle = \langle N_h \rangle + \sum_R p_{R\to h} \langle N_R \rangle = \sum_r p_{r\to h} \langle N_r \rangle\, .
\end{equation}
In order to streamline the formalism, the first sum (over capital $R$) includes only unstable resonances, but the second one sums over all species. 
We generalise $p_{h'\to h} = \delta_{h'h}$, where both $h'$ and $h$ refer to stable hadron species. 
In terms of number densities $n_r(T,\left \{ \mu_i(T)\right \} )$
and the volume we write 
\begin{equation}
\langle N_h^{\mathrm{eff}} \rangle = V(T) \sum_r p_{r\to h} n_r(T,\left \{ \mu_i(T)\right \} )\, .
\end{equation}
It is preferable to use the density as it can be calculated from the temperature $T$ and the set of chemical potentials $\{ \mu_i(T)\}$, where $i$ numbers all stable hadron species. 
Since the effective numbers must stay constant in a cooling system, $d\langle N_h^{\mathrm{eff}} \rangle/dT=0$, and we obtain
\begin{equation}
- \frac{\frac{dV}{dT}}{V(T)} \sum_r p_{r\to h} n_r(T) = \sum_r p_{r\to h} \frac{dn_r}{dT}\, .
\label{e:Vnevol}
\end{equation}
The chemical potentials will be determined from the evolution of $n_r(T)$ in temperature.  However, to do this we would need the term 
$\frac{dV}{dT}/V$. To get this, conservation of total entropy $S$ is assumed: $0=dS/dT = d(sV)/dT$, where $s$ is the entropy density. 
This leads to 
\begin{equation}
\label{e:Vevol}
- \frac{\frac{dV}{dT}}{V} = \frac{\frac{ds}{dT}}{s}\, .
\end{equation}
Inserting into eq.~(\ref{e:Vnevol}) leads to the evolution equation for the number densities
\begin{equation}
\frac{\sum_r p_{r\to h} \frac{dn_r}{dT}}{\frac{ds}{dT}} = \frac{1}{s} \sum_r  p_{r\to h}  n_r(T,\{ \mu_i(T)\})\,  .
\label{e:nevol}
\end{equation}
It is important that $s$ is independent from the densities and can be calculated from $T$ and $\mu_i$'s. Thus the set of equations (\ref{e:nevol})
for each $h$ can be used to evolve the densities and ultimately the chemical potentials. One starts at the chemical freeze-out with equilibrium values 
of the chemical potentials and proceeds to lower temperatures. In general, the chemical potentials increase in a cooling system. 


\section{Results}
\label{s:results}

The yields of resonances are calculated similarly to those of stable hadrons.
We also include higher states which decay into the resonance of interest.
Moreover, since we always look at ratios of abundances, it is adequate to calculate just the ratios of their number densities. 

\begin{figure}[t]
\centering
\includegraphics[width=0.8\textwidth]{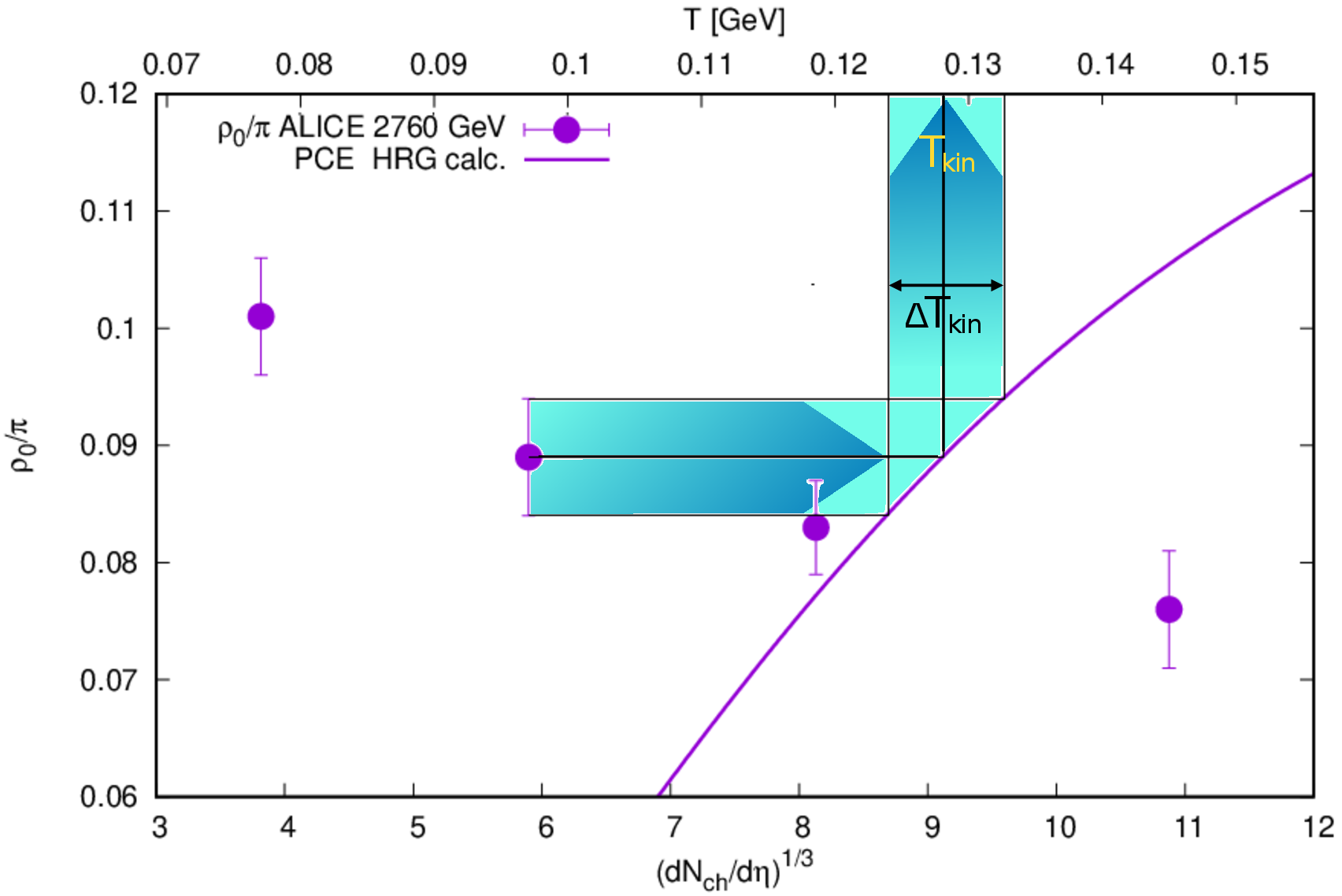}
\caption{The ratio of $2\rho^0/(\pi^++\pi^-)$ from Pb+Pb collisions at $\sqrt{s_{NN}}=2.76$~TeV \cite{ALICE:2018qdv} 
as a function of centrality, 
which is parametrised with the help of the 
charged particle multiplicity as $(dN_{ch}/d\eta)^{1/3}$. Most central collisions correspond to the highest values 
of $(dN_{ch}/d\eta)^{1/3}$.
The theoretical result is calculated as function of temperature, 
which is shown on the upper horizontal axis. To read off the temperature that corresponds to a measured data point, 
the data are horizontally projected onto the theoretical curve and the temperature is read off from the upper axis. 
}
\label{f:rhopi}   
\end{figure}

We present the results in a way explained in Fig.~\ref{f:rhopi} on the example of the $\rho^0/\pi$ ratio. 
%
%
Experimental data \cite{ALICE:2018qdv} 
can be interpreted as coming from PCE with temperatures that can be extracted via comparison
with the theoretical result \cite{Lokos:2022jze}.
For these data, ratios measured in more central collisions indicate lower temperature at which they have been fixed. 
Such a result is consistent with the picture that in central collisions a large system is created that stays together longer than in non-central collisions. 

\begin{figure}[t!]
\centering
\includegraphics[width=0.8\textwidth]{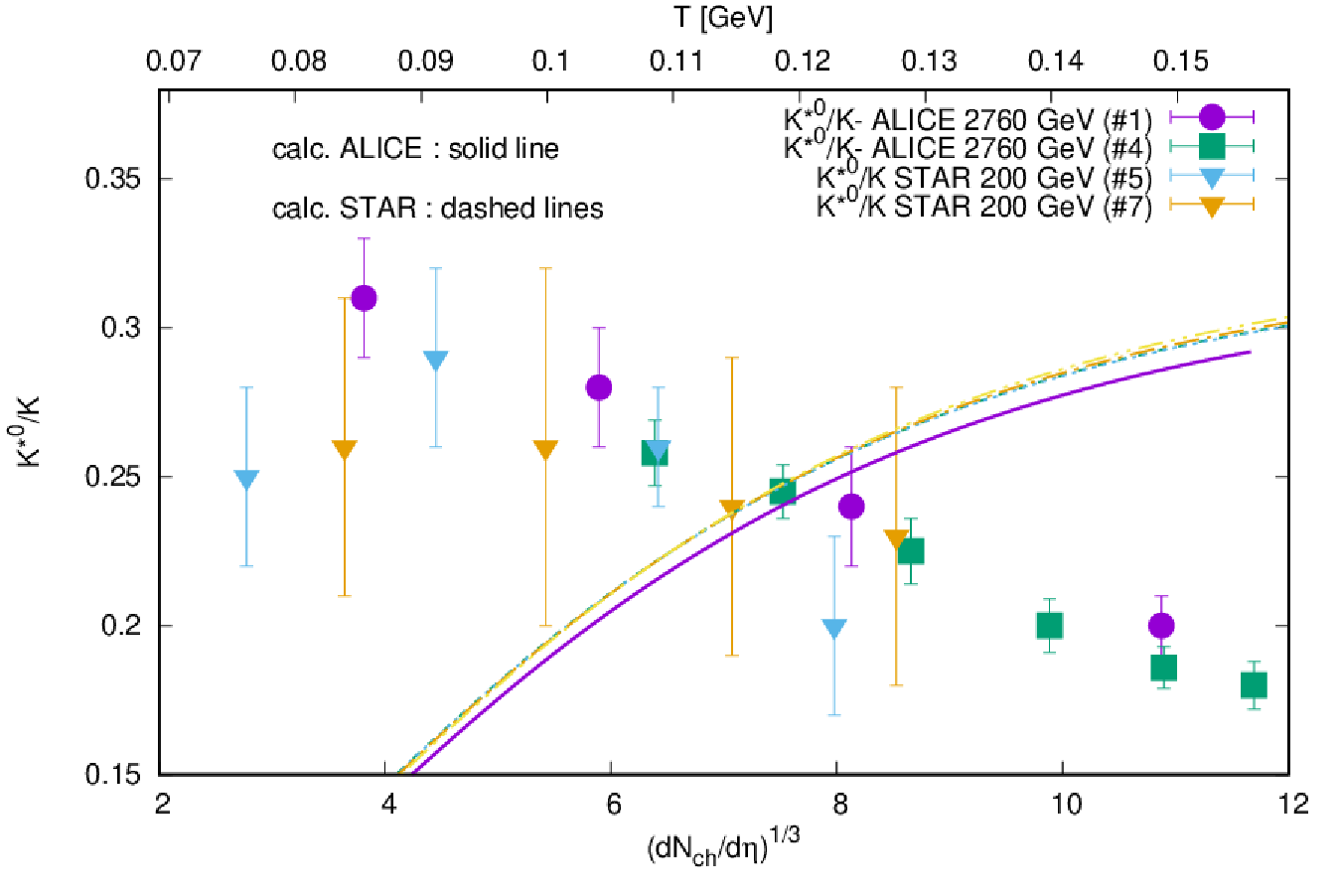}
\caption{%
Ratio of $K^0(892)/K^-$ as function of centrality from Au+Au collisions at $\sqrt{s_{NN}}=200$~GeV 
\cite{STAR:2006vhb,STAR:2010avo}
and  Pb+Pb collisions at $\sqrt{s_{NN}}=2.76$~TeV \cite{ALICE:2014jbq,ALICE:2017ban}. Different (but very close) theoretical 
curves for the Au+Au collisions correspond to different centralities. 
}
\label{f:KsK}       
\end{figure}

Figure~\ref{f:KsK} summarises centrality dependence of the ratio $K^{*0}/K$ from STAR 
\cite{STAR:2006vhb,STAR:2010avo} and ALICE \cite{ALICE:2014jbq,ALICE:2017ban} together with the theoretical predictions. 
%
%
Again, the feature of decreasing temperature with increasing centrality is observed. 

\begin{figure}[h!]
\centering
\includegraphics[width=0.8\textwidth]{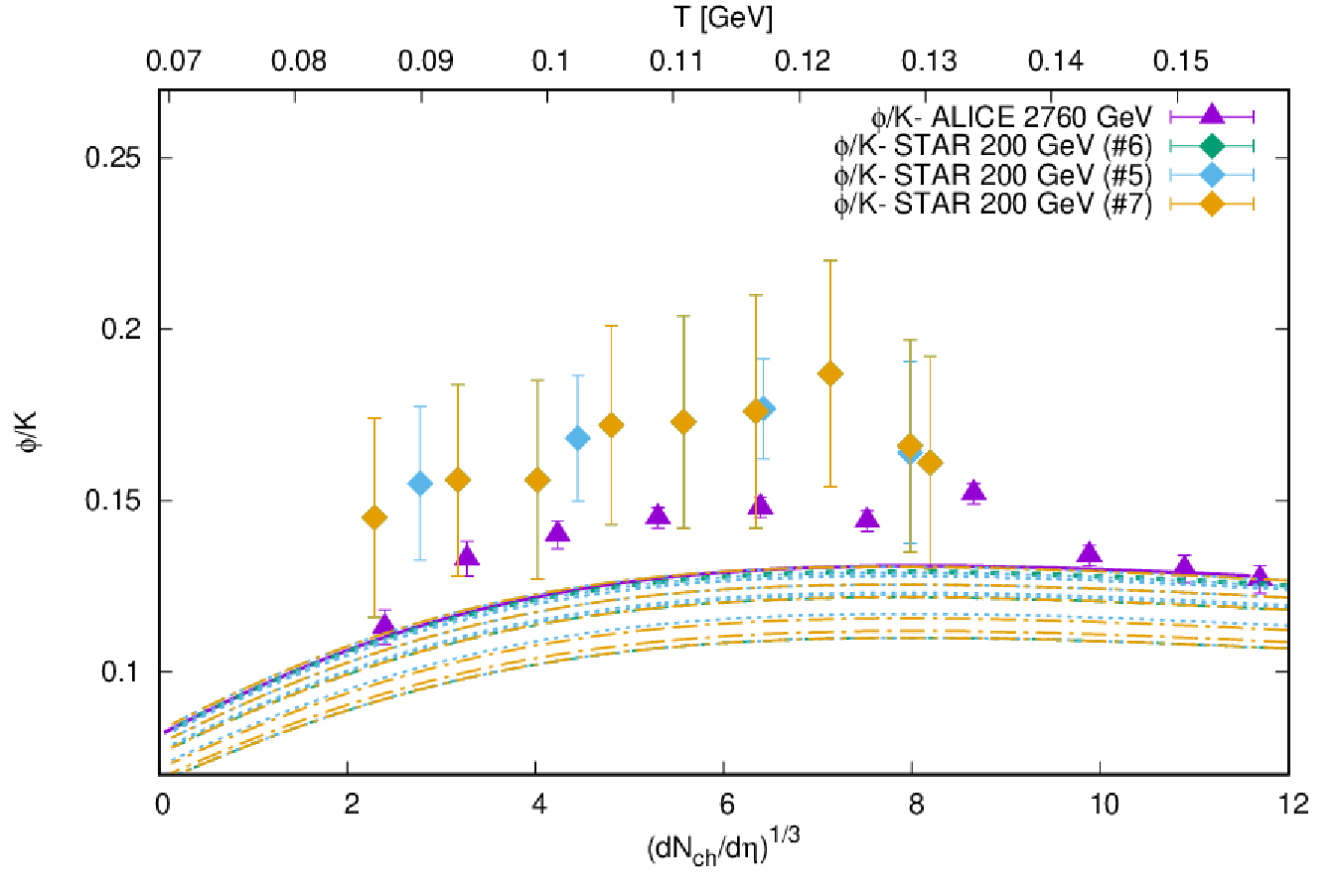}
\caption{%
Ratios of $\phi/K^-$ from Pb+Pb collisions at $\sqrt{s_{NN}}=2.76$~TeV \cite{ALICE:2014jbq} 
and Au+Au collisions  at $\sqrt{s_{NN}}=200$~GeV 
\cite{STAR:2006vhb,STAR:2010avo,STAR:2008bgi}.
Theoretical curves show the temperature dependence of the ratio. For the RHIC energy, 
different curves show the dependence for different centralities. 
}
\label{f:phiK}       
\end{figure}

In Fig.~\ref{f:phiK}
%
%
we present data on $\phi/K^-$ ratio from Au+Au collisions at $\sqrt{s_{NN}}=200$~GeV as measured by 
STAR \cite{STAR:2006vhb,STAR:2010avo,STAR:2008bgi} and from Pb+Pb collisions at 
$\sqrt{s_{NN}} = 2.76$~GeV by the ALICE collaboration \cite{ALICE:2014jbq}.
Again, they are compared with theoretical calculations. In most cases, the measured ratios are clearly above the calculated ones, 
except for the most central collisions at the LHC. 
This may indicate that the $\phi$-meson survives in the hadronic system and does not equilibrate with its decay products. 

\begin{figure}[t!]
\centering
\includegraphics[angle=270,width=0.8\textwidth]{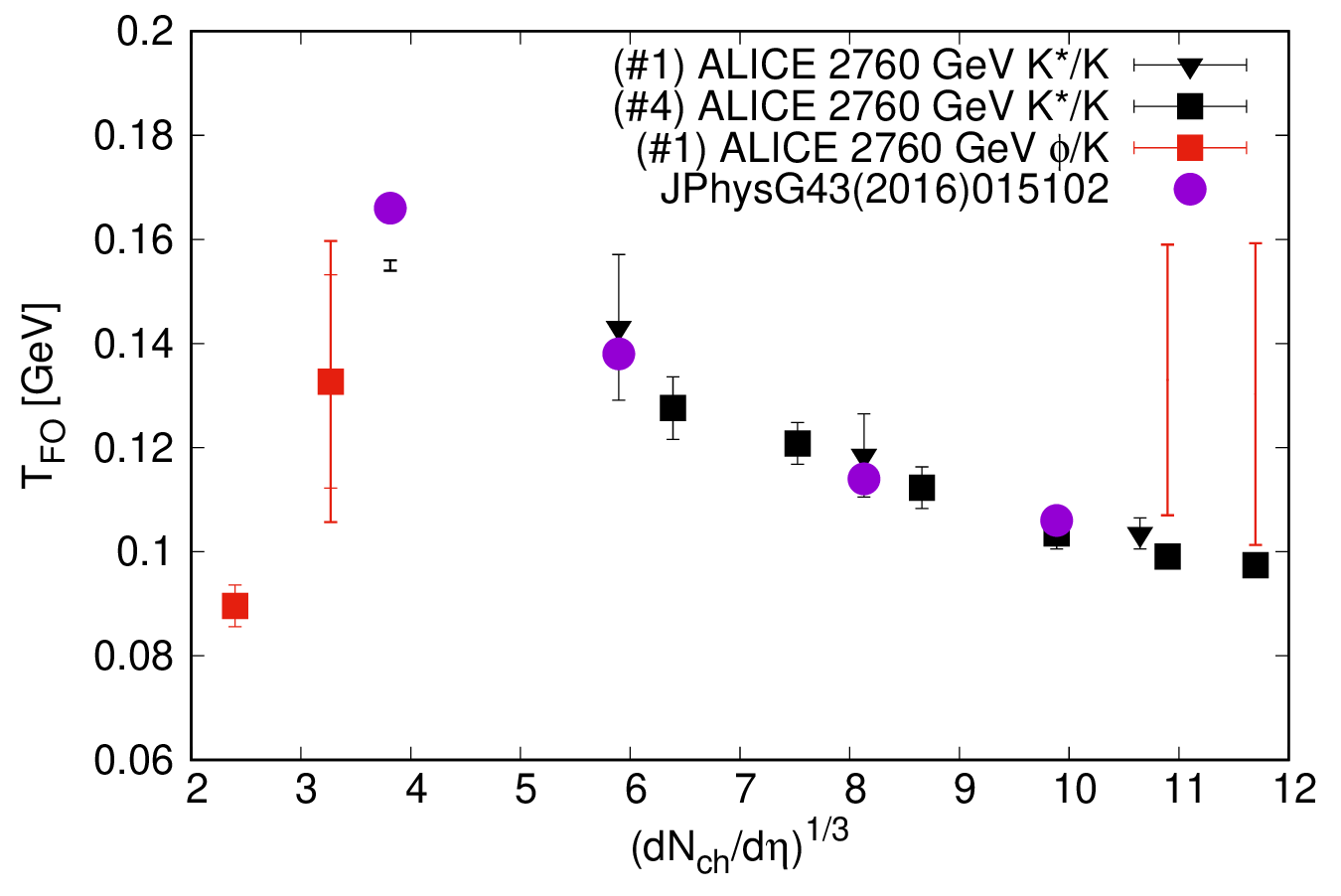}
\caption{
Temperatures obtained from the $K^{*0}/K^-$ \cite{ALICE:2014jbq,ALICE:2017ban} and $\phi/K^-$ \cite{ALICE:2014jbq} 
as function of centrality. Centrality is expressed by $(dN_{ch}/d\eta)^{1/3}$, with most central collisions corresponding 
to its highest  values. The data are superimposed with temperatures obtained from fitting identified single-particle
$p_t$ spectra with blast-wave model \cite{Melo:2015wpa}.
}
\label{f:Ksum}       
\end{figure}

\begin{figure}[t!]
\centering
\includegraphics[width=0.8\textwidth]{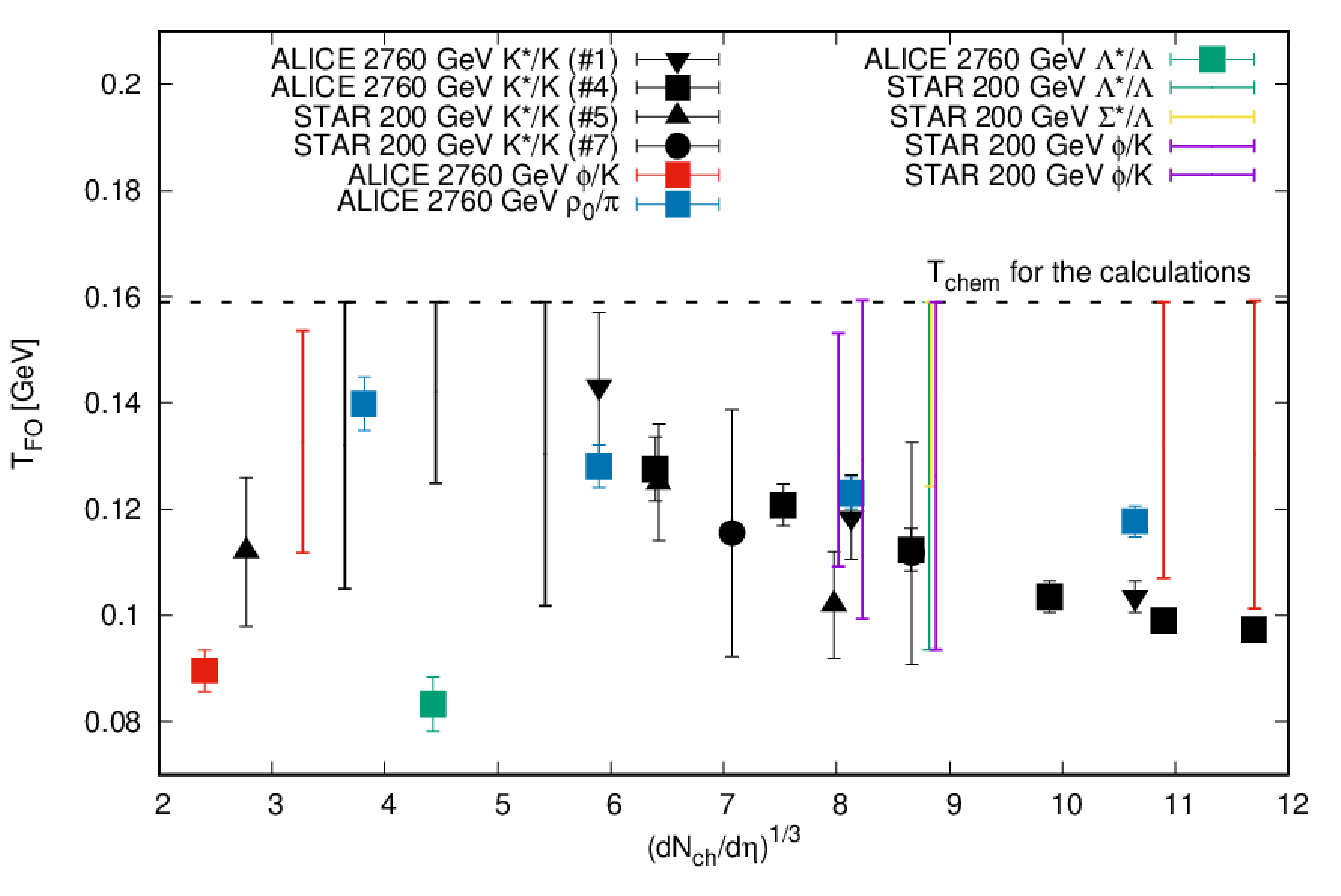}
\caption{
Summary of all temperatures depending on centrality (expressed via $(dN_{ch}/d\eta)^{1/3}$).
The temperatures are extracted from ratios of resonances. If only uncertainty intervals are shown without 
a data point, then the experimental result could not fit on the theoretical curve, but its  uncertainty interval had some 
overlap with it. Data from ALICE on $K^{*0}/K^-$ \cite{ALICE:2014jbq,ALICE:2017ban}, 
$\rho^0/\pi$ \cite{ALICE:2018qdv}, $\phi/K$ \cite{ALICE:2014jbq}, $\Lambda^*/\Lambda$ \cite{ALICE:2018ewo}, 
and from STAR on $K^{*0}/K^-$ \cite{STAR:2006vhb,STAR:2010avo}, 
$\Lambda^*/\Lambda$ \cite{STAR:2006vhb}, $\Sigma^*/\Lambda$ \cite{STAR:2006vhb}, 
and $\phi/K^-$ \cite{STAR:2006vhb,STAR:2010avo}. 
Horizontal dashed line indicates the temperature of the chemical freeze-out, which is the maximum 
temperature in this model.
}
\label{f:tsum}       
\end{figure}

The temperatures that we obtained from the $K^{*0}/K^-$ and $\phi/K^-$ ratios are compared in Fig.~\ref{f:Ksum} 
%
%
with values inferred from fits with the blast-wave model to the identified single-particle $p_t$ spectra \cite{Melo:2015wpa}.
While from $\phi/K^-$ barely any reasonable temperatures can be obtained, the values from $K^{*0}/K^-$
coincide with the results from fitting of the $p_t$ spectra. 
We also provide a comprehensive summary of all extracted temperatures in Fig.~\ref{f:tsum}.
%
%
While the ratios of $\rho^0/\pi$ and $K^{*0}/K$ fit into the overall picture with $T_{\mathrm FO}$ that decreases with 
increasing centrality of the collisions, the ratios of $\phi/K$ and those involving excited hyperons either 
largely depart from the common centrality dependence or even cannot be interpreted by any temperature within 
the PCE model.


\section{Conclusions}
\label{s:conc}

We have shown \cite{Lokos:2022jze} that while PCE can nicely interpret centrality dependence of 
$\rho^0/\pi$ and $K^{*0}/K^-$ ratios, it cannot reproduce $\phi/K^-$. The $\phi$ mesons appear to be too copious 
in comparison with the model. 

The model relies on several simplifying assumptions which may be revisited. It assumes isentropic expansion
in order to represent the growth of the volume through the decrease of the entropy density. 
Also,  resonances are accounted for as narrow stable particles. This treatment can be improved by explicitly 
including the interactions via phase shifts.


\acknowledgments
BT is supported by the Czech Science Foundation (GA\v{C}R) under No 22-25026S. 
BT also acknowledges the support by VEGA  1/0521/22.
SL is grateful of the support of Hungarian National E\"otv\"os Grant established by the Hungarian Government.


\begin{thebibliography}{99}

\bibitem{Andronic:2017pug}
\BY{Andronic A., Braun-Munzinger P., Redlich K. \atque Stachel J.}
\IN{Nature}{561} {2018}{321}

\bibitem{Andronic:2016nof}
\BY{Andronic A., Braun-Munzinger P., Redlich K. \atque Stachel J.}
\IN{J. Phys. Conf. Ser.}{779} {2017}{012012}

\bibitem{Dashen:1969ep}
\BY{Dashen R., Ma S.K. \atque Bernstein H.J.}
\IN{Phys. Rev.}{187} {1969}{345}

\bibitem{Venugopalan:1992hy}
\BY{Venugopalan R. \atque Prakash M.}
\IN{Nucl. Phys. A}{546}{1992}{718}

\bibitem{Andronic:2018qqt}
\BY{Andronic A.,  Braun-Munzinger P., Friman B., Lo P.~M., Redlich K. \atque Stachel J.}
\IN{Phys. Lett. B}{792}{2019}{304}

\bibitem{Mazeliauskas:2019ifr}
\BY{Mazeliauskas A. \atque Vislavicius V.}
\IN{Phys. Rev. C}{101}{2020}{014910}

\bibitem{Melo:2019mpn} 
\BY{Melo I. \atque Tom\'a\v{s}ik B.} 
\IN{J. Phys. G}{47} {2020}{045107} 

\bibitem{Bebie:1991ij}
\BY{Bebie H., Gerber P., Goity J.L. \atque Leutwyler H}
\IN{Nucl. Phys. B}{378}{1992}{95}

\bibitem{Lokos:2022jze}
\BY{L\"ok\"os S. \atque Tom\'a\v{s}ik B.}
\IN{Phys. Rev. C}{106}{2022}{034912}

\bibitem{ALICE:2018qdv}
\BY{Acharya S. \textit{et al.} [ALICE]}
\IN{Phys. Rev. C}{99}{064901}{2019}

\bibitem{STAR:2006vhb}
\BY{Abelev B.I. \textit{et al.} [STAR]}
\IN{Phys. Rev. Lett.}{97}{2006}{132301}

\bibitem{STAR:2010avo}
\BY{Aggarwal M.M. \textit{et al.} [STAR]}
\IN{Phys. Rev. C}{84}{2011}{034909}

\bibitem{ALICE:2014jbq}
\BY{Abelev B.B. \textit{et al.} [ALICE]}
\IN{Phys. Rev. C}{91}{2015}{024609}

\bibitem{ALICE:2017ban}
\BY{Adam J. \textit{et al.} [ALICE]}
\IN{Phys. Rev. C}{95}{2017}{064606}

\bibitem{STAR:2008bgi}
\BY{Abelev B.I. \textit{et al.} [STAR]},
\IN{Phys. Rev. C}{79}{2009}{064903}

\bibitem{Melo:2015wpa}
\BY{Melo I. \atque Tom\'a\v{s}ik B.}
\IN{J. Phys. G}{43}{2016}{015102}



\bibitem{ALICE:2018ewo}
\BY{Acharya S. \textit{et al.} [ALICE]}
\IN{Phys. Rev. C}{99}{2019}{024905}


\end{thebibliography}
\end{document}